\documentclass[aps,pre,twocolumn,superscriptaddress,amsmath,amssymb,floatfix,showkeys]{revtex4}
\usepackage{graphicx}
\usepackage{bigints}
\usepackage{times}
\begin{document}
\def\D{\Delta}
\def\d{\delta}
\def\r{\rho}
\def\p{\pi}
\def\a{\alpha}
\def\g{\gamma}
\def\ra{\rightarrow}
\def\s{\sigma}
\def\b{\beta}
\def\e{\epsilon}
\def\G{\Gamma}
\def\om{\omega}
\def\l{\lambda}
\def\f{\phi}
\def\w{\psi}
\def\m{\mu}
\def\t{\tau}
\def\c{\chi}
 \title{Chaos and Unpredictability in Evolution}

\author{Iaroslav Ispolatov}
\email{jaros007@gmail.com}
\affiliation{
Departamento de Fisica, Universidad de Santiago de Chile,
Casilla 302, Correo 2, Santiago, Chile}
\author{Michael Doebeli}
\email{doebeli@zoology.ubc.ca}
\affiliation{Department of Zoology and Department of
  Mathematics,  University of British Columbia, 6270 University Boulevard, Vancouver B.C. Canada, V6T 1Z4}

\begin{abstract}The possibility of complicated dynamic behaviour driven by non-linear feedbacks in dynamical systems has revolutionized science in the latter part of the last century. Yet despite examples of complicated frequency dynamics, the possibility of long-term evolutionary chaos is rarely considered. The concept of ``survival of the fittest'' is central to much evolutionary thinking and embodies a perspective of evolution as a directional optimization process exhibiting simple, predictable dynamics. This perspective is adequate for simple scenarios, when frequency-independent selection acts on scalar phenotypes. However, in most organisms many phenotypic properties combine in complicated ways to determine ecological interactions, and hence frequency-dependent selection. Therefore, it is natural to consider models for the evolutionary dynamics generated by frequency-dependent selection acting simultaneously on many different phenotypes. Here we show that complicated, chaotic dynamics of long-term evolutionary trajectories in phenotype space is very common in a large class of such models when the dimension of phenotype space is large, and when there are epistatic interactions between the phenotypic components. Our results suggest that the perspective of evolution as a process with simple, predictable dynamics covers only a small fragment of long-term evolution. Our analysis may also be the first systematic study of the occurrence of chaos in multidimensional and generally dissipative systems as a function of the dimensionality of phase space. 

\end {abstract}


\maketitle
\section{Author Summary}

40 years ago, the discovery of deterministic chaos has revolutionized science. Surprisingly, few of these insights have entered the realm of evolutionary biology, where “survival of the fittest” epitomizes evolution as an optimization process that generally converges to an equilibrium, the optimal phenotype. This perspective may be correct for simple phenotypes, such as body size, but in reality, all organisms have a multitude of phenotypic properties that impinge on birth and death rates, and hence on evolutionary dynamics. However, evolution in high-dimensional phenotype spaces is rarely studied due to the formidable technical difficulties involved. 
	In the enclosed paper, we have used the recently developed mathematical framework of adaptive dynamics for a systematic  investigation of long-term evolutionary dynamics in high-dimensional phenotype spaces. Our main conclusions are that chaotic evolution is common in complicated phenotype spaces. This is relevant for Gould's famous question about ``replaying the tape of life'': if evolution is fundamentally chaotic, then evolution is generally unpredictable in the long term, even if selection is deterministic. Our results show that evolutionary chaos is indeed common, and hence unpredictability is the rule rather than the exception. This suggests that the perspective of evolution as an equilibrium process must be fundamentally revised.

\section{Introduction}

Evolution generally takes place in complex ecosystems and
is affected by many different processes that generate non-linear
dependencies. According to general dynamical systems theory, which has
shown that even simple dynamical system can exhibit complicated
dynamics   \cite{li_yorke1975,may1976,bak_etal1987,gleick1988}, one would therefore expect that evolutionary dynamics tend to be complicated. However, this is contrary to traditional evolutionary thinking, which is based on the concept of ``survival of the fittest'', and on metaphors of static fitness landscapes  \cite{wright1932,gavrilets2004,svensson_calsbeek2012}, in which evolution optimizes simple, scalar phenotypes such as body size, age and size at maturity, fecundity, stress tolerance, antibiotic resistance, etc. Accordingly, the ``fittest'' type wins, and hence evolution is often envisioned as a dynamical system that converges to an equilibrium in phenotype space, representing the optimally adapted type. It is of course generally acknowledged that over large time scales, evolution is a non-stationary process, but this is usually attributed to long-term changes in the external environment causing shifts in evolutionary optima.

Static fitness landscapes describe frequency-independent selection whose strength and direction is not affected by the current phenotypic composition of an evolving population. However, it is widely recognized that ecological interactions, such as competition and predation, often lead to frequency-dependent selection, in which the current phenotypic composition of a population determines whether a particular phenotype is advantageous or not  \cite{heino_etal1998,schluter2000,doebeli2011}. For example, whether it is advantageous to have a preference for a particular type of food depends on the preferences of the other individuals in the population. Frequency dependence generates an evolutionary feedback loop, because selection pressures, which cause evolutionary change, change themselves as a population's phenotype distribution evolves. It is well-known that this feedback can produce complicated dynamics in models in which the dynamic variables are the frequencies of a fixed and finite set of different types in a given population \cite{altenberg1991, gavrilets_hastings1995, nowak_sigmund1993, schneider2008, priklopil2012}.
However, such models are essentially ecological models, in which chaotic dynamics is the result of coexistence of different types due to frequency-dependent selection. They are ecological models because they describe the dynamics of the (relative) abundance of different types on short, ecological time scales. In particular, the phenotypes or genotypes present in the population never go beyond the finite set initially provided by the model. Perhaps this essentially ecological nature of these models helps explain why, even though such models have been shown to exhibit complicated dynamics, the possibility of evolutionary chaos does not really play a role in mainstream evolutionary thinking.

It is important to distinguish models for short-term frequency dynamics from evolutionary models in which the dynamic variables are the (mean) phenotypes themselves, and which track the trajectories of such phenotypes in continuous phenotype spaces and over long evolutionary time scales. The phase space for this type of model is the space of all possible phenotypes (rather than the space of frequencies of different types). Adaptive dynamics \cite{geritz_etal1998, dieckmann_law1996} provides a useful framework for generating models of long-term evolutionary dynamics of phenotypes. Intuitively, adaptive dynamics unfolds as a series of phenotypic substitutions that give rise to evolutionary trajectories in phenotype space. Frequency dependence plays an important role in adaptive dynamics, but most often, this feedback mechanism has been studied in relatively simple scenarios,
in which frequency dependence is still generally expected to generate long-term equilibrium dynamics. 
But even in simple phenotype spaces, frequency dependence can lead to interesting evolutionary phenomena, such as adaptive diversification  \cite{doebeli2011,geritz_etal1998,doebeli_dieckmann2000}. In more complicated phenotype spaces containing scalar phenotypes of each
of a number of co-evolving populations, frequency dependence can generate complicated evolutionary dynamics. For example, coevolution of scalar traits in predator and prey populations can lead to arms races in the form of cyclic dynamics in phenotype space \cite{dieckmann_etal1995,dieckmann_law1996,dercole_etal2006}, and coevolution of scalar traits in a three-species food chain can generate chaotic dynamics in phenotype space \cite{dercole_rinaldi2008}. In all these examples, the dynamic variables undergoing evolution are the (mean) traits in the various interacting species, and the trajectories of these traits in combined phenotype space comprising the scalar traits of all interacting species can exhibit complicated dynamics. 

Long-term evolutionary dynamics of continuous phenotypes is ultimately driven by birth and death rates of individual organisms. In general, these birth and death rates are determined in a complicated way by many different phenotypic properties, which could be as diverse as the molecular efficiency of photosynthesis and the height of trees. Therefore, even for single species it is natural to study evolutionary dynamics 
in high-dimensional phenotype spaces. For frequency-independent selection, evolution is still an optimization process in such spaces, although genetic correlations between phenotypic components may warp the fitness landscape and alter the convergence dynamics to the local optima  \cite{lande1979}. Yet,  apart from a few examples \cite{doebeli_ispolatov2010,tucker_etal2012}, little is known about the expected complexity of the long-term evolutionary dynamics of high-dimensional phenotypes when selection is frequency-dependent. Here we ask whether frequency dependence due to competition in high-dimensional phenotype spaces of  single species yields evolutionary dynamics that are fundamentally different from the equilibrium dynamics resulting from evolution in simple phenotype spaces.

\section{Model and Results}
We use adaptive dynamics theory  \cite{geritz_etal1998,dieckmann_law1996} to study the long-term evolutionary dynamics in a large class of multidimensional single-species competition models. The starting point is the  widely used logistic model   \cite{doebeli2011}
\begin{equation}
\label{lm}
\frac{\partial N(x,t )}{\partial t}=rN(x,t) \left(1-\frac{\bigintsss \a(x,y)
 N(y,t) dy }{K(x) }\right).
\end{equation}
Here $N(x,t)$ is the density of individuals of phenotype $x$ at time $t$, and $K(x)$ is the carrying capacity of a monomorphic population consisting entirely of $x$-individuals. The competitive
impact between individuals of phenotypes $x$ and $y$ is given by the
competition kernel $\alpha(x,y)$, so that an $x$-individual experiences an effective density $\bigintssss \a(x,y)N(y,t) dy$. This model has been used extensively to study the evolutionary dynamics of scalar traits $x\in  \mathbb{R}$  \cite{doebeli2011}. Here we assume more generally that $x\in \mathbb{R}^d$ is a $d$-dimensional vector describing $d\geq 1$ scalar phenotypic properties. We also assume that $\a(x,x)=1$ for all $x$, and that
the intrinsic growth rate $r$ is independent of the phenotype $x$ and is equal to 1. To derive the adaptive dynamics of the multidimensional trait $x$, we consider a resident population that is monomorphic for trait $x$, for which the ecological model (1) has a unique, globally stable equilibrium density $K(x)$, regardless of the dimension $d$ of $x$. Assuming that the resident is at its ecological equilibrium $K(x)$, the invasion fitness $f(x,y)$ of a rare mutant $y$ is its per capita growth rate in the resident population $x$,
\begin{equation}
\label{if}
f(x,y)=1-\frac{\alpha(y,x)K(x)}{K(y)}.
\end{equation}
The selection gradient $s(x)=(s_1(x),\dots,s_d(x))$ is derived from the invasion fitness as 
\begin{equation}
\label{sg}
s_i(x)=\left.\frac{\partial f(y,x)}{\partial y_i}\right\vert_{y=x}=
- \left.\frac{\partial \a(y,x)}{\partial y_i}\right\vert_{y=x} + \frac{\partial K(x)}{\partial x_i}\frac{1}{K(x)}.
\end{equation}
Finally, the adaptive dynamics of the trait $x$ is
\begin{equation}
\label{ad}
\frac{dx}{dt}=M(x)\cdot s(x),
\end{equation}
where $M(x)$ is a $d\times d$-matrix describing the mutational process in the $d$ phenotypic components  \cite{leimar2009,doebeli2011} (and where $dx/dt$ and $s(x)$ are column vectors). In general, the entries of $M(x)$ depend on the current population size, and hence implicitly on $x$, but for simplicity we assume here that $M(x)$ is the identity matrix, which is a conservative assumption as far as the complexity of the adaptive dynamics (4) is concerned.

Complicated dynamics in the form of oscillations can already occur if the selection gradient $s(x)$ in (4) is linear. In fact, with randomly chosen coefficients, the probability of oscillatory behaviour is $1-2^{-d(d-1)/4}$, and hence rapidly approaches 1 as the dimension of phenotype space is increased  \cite{edelman1997}. To study non-linear systems, we assume that the complexity of the interactions between phenotypic components in determining ecological properties is contained in the competition kernel $\a(x,y):\mathbb{R}^d\times \mathbb{R}^d\rightarrow \mathbb{R}$, which is in general a complicated, nonlinear function that reflects epistatic interactions between the different phenotypic components. We only consider the Taylor expansion of this function up to quadratic terms, and absorbing constant terms into a change of coordinates, we get 
\begin{equation}
\label{te}
\left.\frac{\partial \a(y,x)}{\partial
    y_i}\right\vert_{y=x} \approx - \sum_{j=1}^d b_{ij} x_j - \sum_{j,k=1}^d
a_{ijk}x_j x_k
\end{equation}
For the carrying capacity, we assume a simple symmetric form: 
$K(x)=\exp(-\sum_i x_i^4/4)$, which, together with the competition
kernel gradient (\ref{te}),  ensures that the trajectories of the adaptive dynamics (4)
are confined to a finite region of phenotype space. 
With these assumptions, the 
adaptive dynamics (4), describing the evolution of the multidimensional phenotype $x$, becomes
\begin{equation}
\label{main}
\frac{dx_i}{dt}=\sum_{j=1}^d b_{ij} x_j + \sum_{j,k=1}^d
a_{ijk}x_j x_k - x_i^3, \; i=1,\ldots,d.
\end{equation}
The parameters $b_{ij}$ and $a_{ijk}$ reflect the epistatic interactions among the $d$ phenotypic components. If $b_{ij}=0$ for $i\neq j$ and $a_{ijk}=0$ for $i\neq j,k$, there is no epistasis between the phenotypic components, and in that case, the adaptive dynamics (6) always converges to an equilibrium in phenotype space. Thus, if system (6) exhibits complicated dynamics, it must be due to epistatic interactions. To address the question of the ubiquity of complex evolutionary dynamics, we chose, for a given dimension $d$ of phenotype space, many different sets of parameters $b_{ij}$ and $a_{ijk}$, reflecting a wide range of possible epistatic interaction structures for the phenotypic components. Specifically, parameters were drawn randomly from a Gaussian
distribution with mean zero and variance 1, and for each choice of parameters evolutionary trajectories were obtained by integrating the adaptive dynamics (6) numerically starting from a random initial condition. We show in section C of the {\it Methods} that the results are qualitatively the same for any distribution of the coefficients $b_{ij}$ and $a_{ijk}$ with a finite variance, and that the results remain true if the coefficients are rescaled such the total strength of epistatic interactions is independent of the dimension $d$ of phenotype space. 
For each
trajectory we measured
the time average of the largest Lyapunov exponent $\l $
 \cite{devaney1986} (see {\it Methods}). 
Based on Lyapunov exponents, we classified the evolutionary trajectories into three groups: fixed points ($\l < - 0.1$), periodic
or quasi-periodic attractors ($| \l | < 0.1$), and chaotic attractors ($ \l  > 0.1$). Examples are shown in Fig.~1.
\begin{figure}
\includegraphics[width=0.45\textwidth]{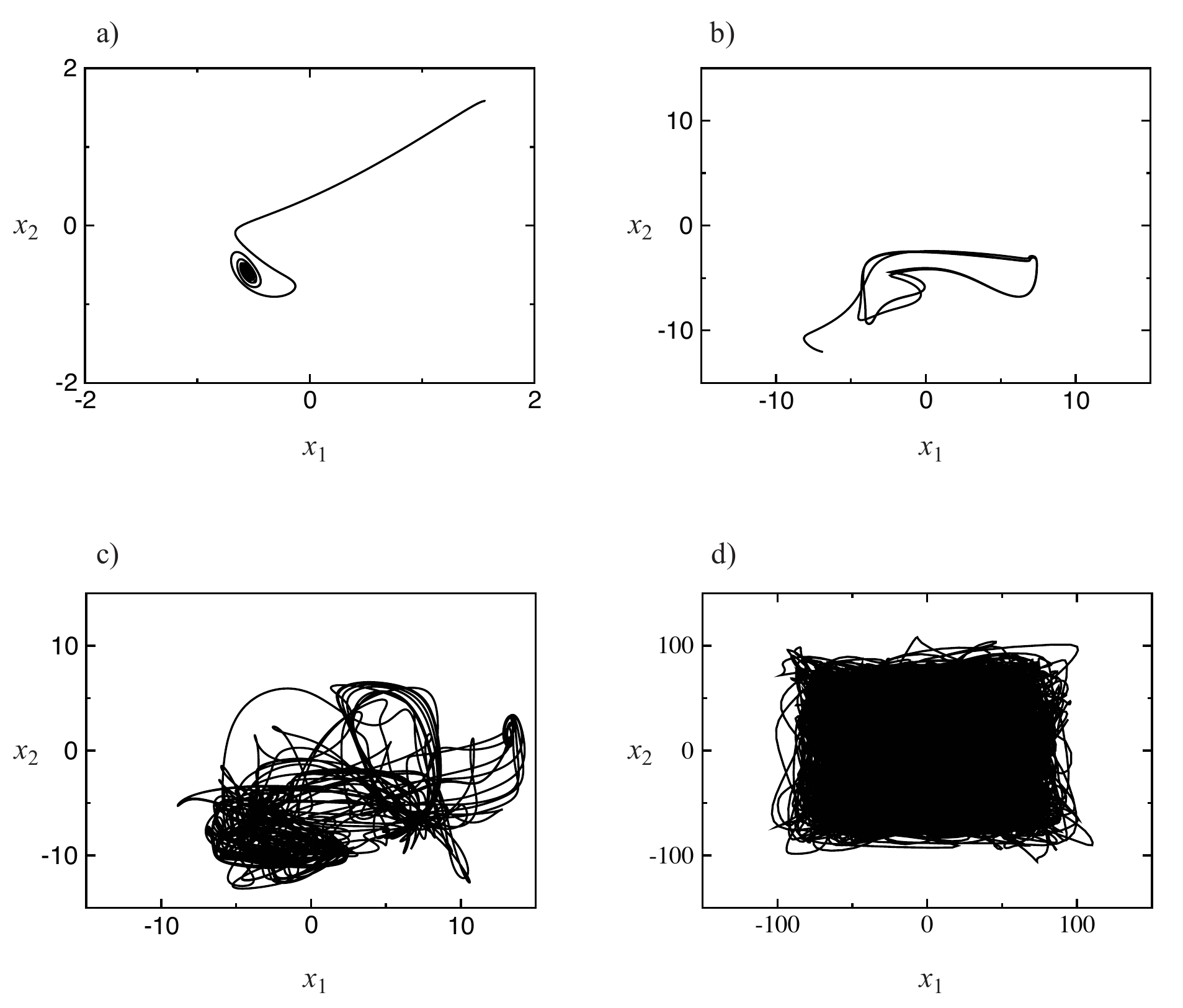}
\caption {\label{f1} 
 a) Example of equilibrium dynamics for $d=5$; largest Lyapunov exponent $\l=-0.12$. 
b) Example of quasi-periodic dynamics for $d=15$; largest Lyapunov exponent $\l=0.008$.
c) Example of chaotic dynamics on a non-ergodic (``strange'') attractor for $d=15$; largest Lyapunov exponent $\l=1.35$.
d) Example of ergodic chaotic dynamics for $d=100$; largest Lyapunov exponent $\l=1850$. The trajectory essentially fills the phenotype space on a scale of $(-d,d)$ in each phenotypic dimension. Here and below the integration of (\ref{main}) was performed from $t=0$ to $t=400/d^2$ using a 4th-order Runge-Kutta method with time step $dt=0.1 d^{-2}$. The coefficients $a_{ij}$ and $b_{ijk}$ were randomly drawn from a Gaussian distribution with zero mean and unit variance, and the initial conditions were randomly drawn from a Gaussian distribution with mean 0 and variance $d^2$. The panels show projections of the evolutionary trajectories onto a randomly chosen 2-dimensional subspace of the phenotype space.
}\end{figure}

\begin{figure}
\includegraphics[width=0.45\textwidth]{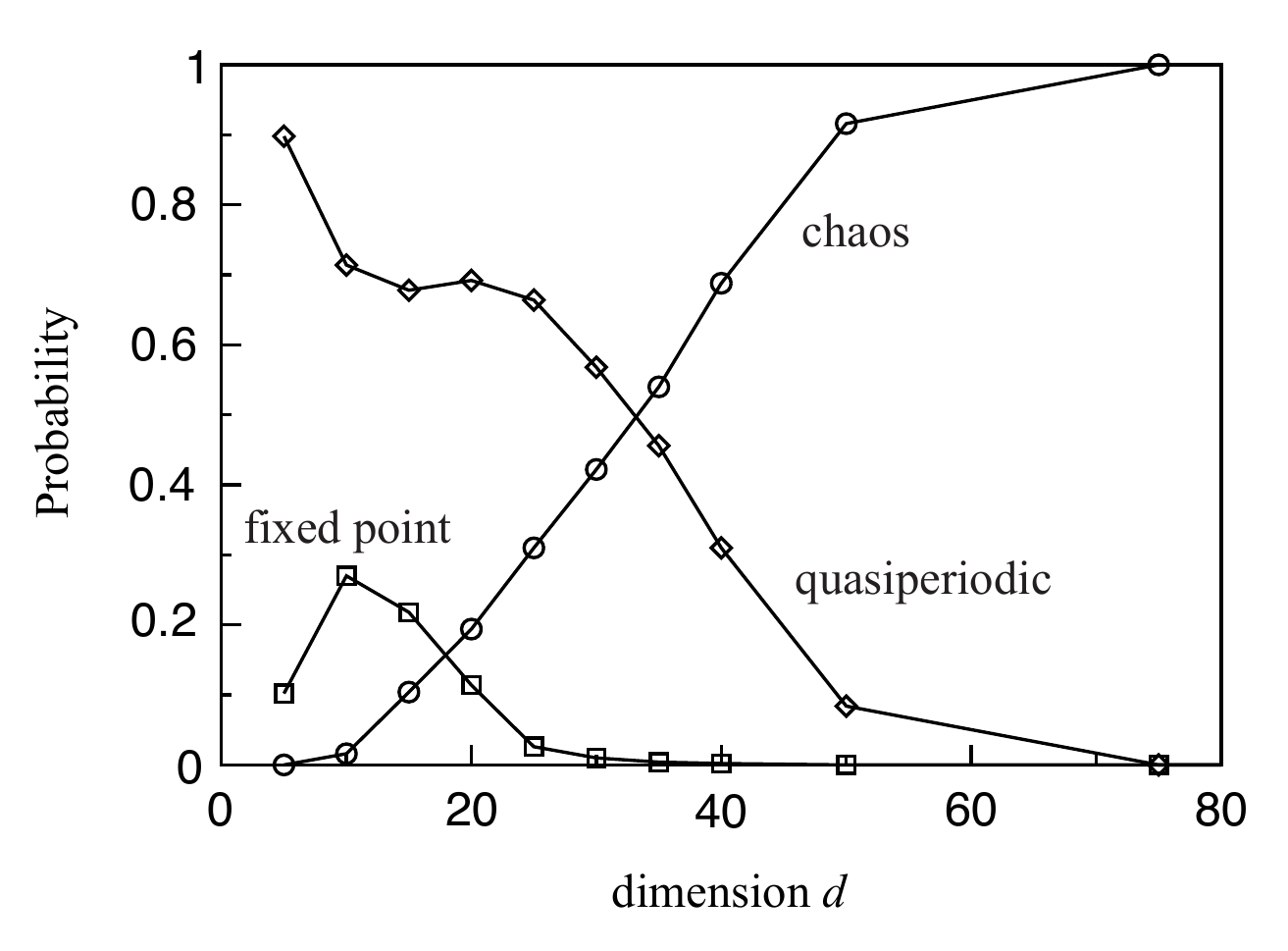}
\caption {\label{f2} 
  Percentage of different types of dynamics as a function of the dimensionality $d$ of phenotype space. For each $d$, we generated 50 instances of the dynamical system (6) by choosing the coefficients $b_{ij}$ and $a_{ijk}$ randomly (see Fig. 1 legend) and then numerically integrating system (6) with 4 sets (single set for $d=200$) of random initial conditions. For each $d$, the figure shows the percentage of instances resulting in equilibrium dynamics with Lyapunov exponent $\l<-0.1$ (squares), quasi-periodic dynamics with Lyapunov exponent $-0.1<\l<0.1$ (diamonds), and chaos with Lyapunov exponent $\l>0.1$ (circles). Chaos starts to occur in a significant fraction of dynamical systems for $d\approx15$, and starts to dominate for $d\approx35$, with essentially no non-chaotic dynamics for $d\geq75$. 
}\end{figure}

Our main result is that the probability of chaos increases
with the dimensionality $d$ of the evolving system, approaching 1  for
$d\sim 75$ (Fig.~2). Moreover, already for $d\gtrsim15$, the majority of chaotic trajectories 
become ergodic  \cite{devaney1986}, and hence essentially fill out the available phenotype space over evolutionary time. The size of the filled phenotype  space  scales approximately as $|x_i|<d$ for
each phenotypic dimension $i$ ({\it Methods}), and the density of trajectories exhibits a 
universal probability distribution (Fig.~3). 
\begin{figure}
\includegraphics[width=0.45\textwidth]{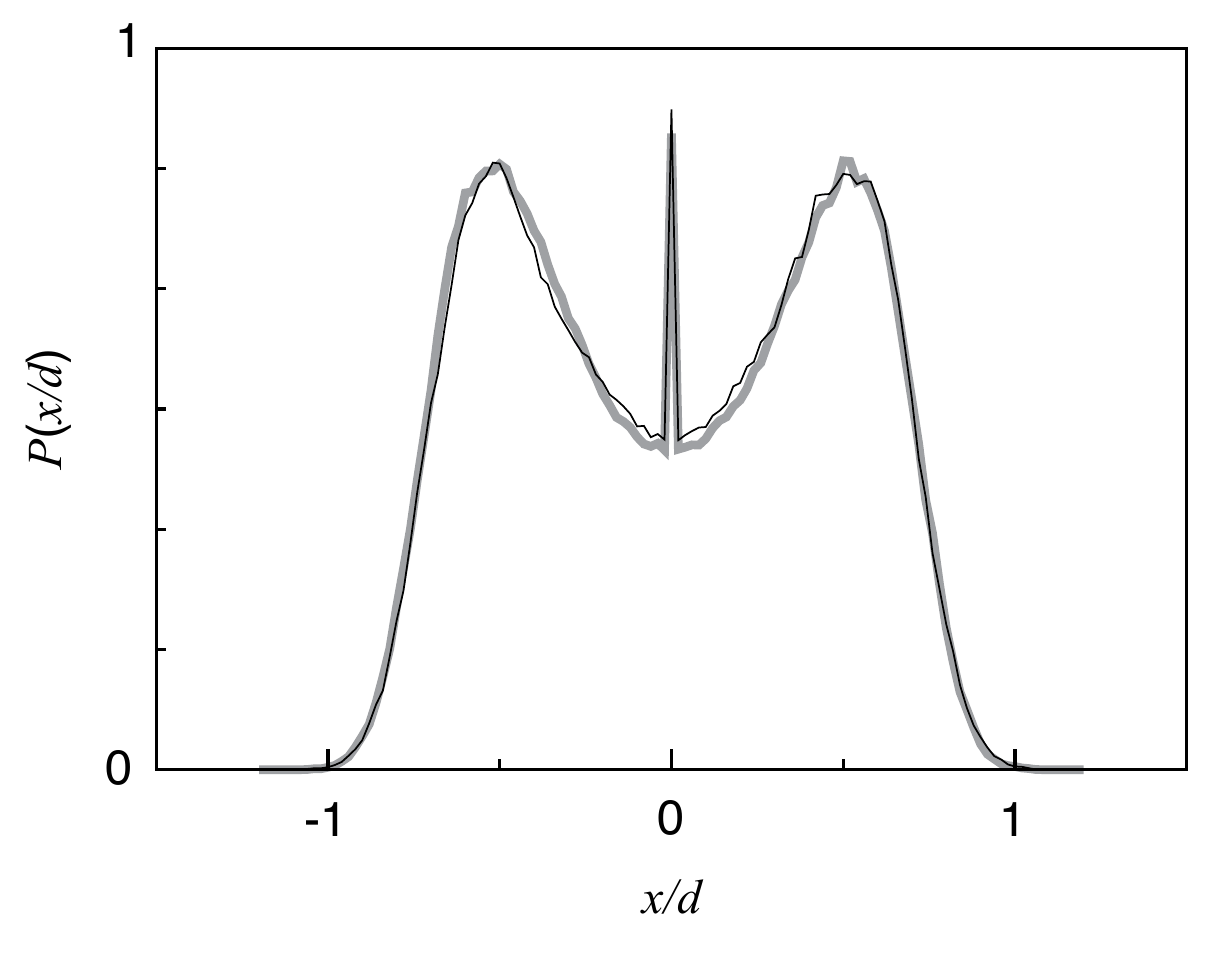}
\caption {\label{f3} 
  Examples of the density distributions $P(x/d)$ for $d=100$ (thick grey line) and $d=150$ (thin black line) of the scaled state variable $x/d$ ({\it Methods}) in an arbitrary component of phenotype space. Once the system is in the ergodic regime, the distribution is universal in the sense that it does not depend on the dimension $d$ of phenotype space, on the particular choice of coefficients in (6), or on the phenotypic component. The distribution is obtained as a histogram over all the states that a system's trajectory attains during a long period of time. It shows two peaks lying symmetrically around the phenotype $x=0$, which corresponds to the maximum of the carrying capacity function $K(x)$, and near which the system also spends much of its time, which is reflected in the third, central peak of the distribution $P(x/d)$. 
}\end{figure}
It is important to note that because these ergodic trajectories fill out large areas of phenotype space, they are very different from noisy equilibrium points. Finally, we observe that the largest Lyapunov exponent converges to
the universal asymptotic $\l \sim d^2$ (Fig. 4). 
\begin{figure}
\includegraphics[width=0.45\textwidth]{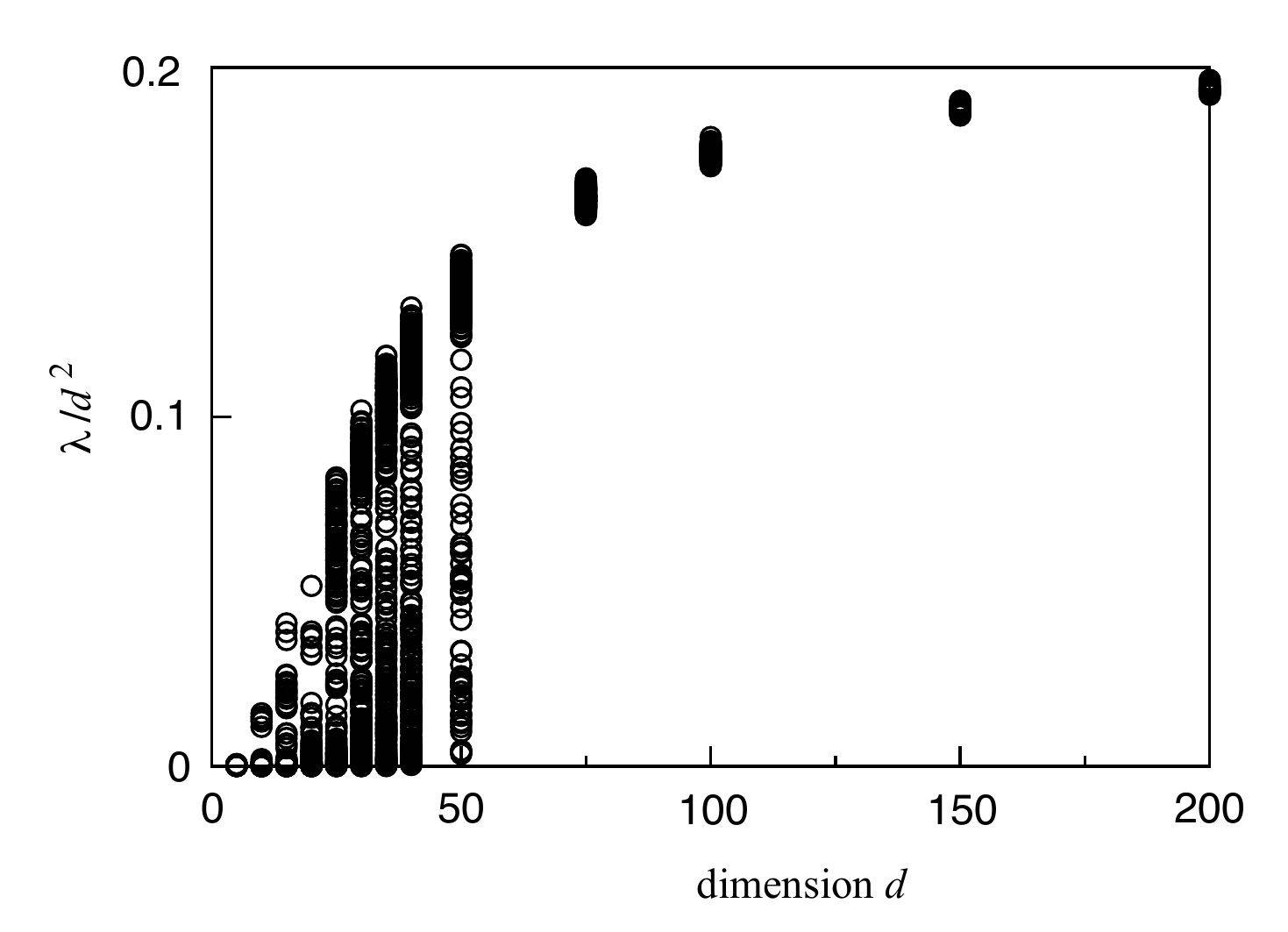}
\caption {\label{f4} 
  The scaled largest Lyapunov exponent $\l/d^2$ ({\it Methods}) as a function of the dimension $d$ of phenotype space, for the same data set as used for Fig.~2. For large $d$, the largest scaled Lyapunov exponent saturates at an asymptotic value $\l^* \approx  0.235$ ({\it Methods}).
}\end{figure}
In {\it Methods} we provide qualitative analytical explanations for these numerical results. In particular, we derive an analytical approximation for the probability of chaos as a function of the dimension of phenotype space (Fig. 9) by arguing that a trajectory is chaotic if all fixed points of system (6) have at least one repelling direction, which becomes certain for large $d$ due to the epistatic interactions between phenotypic components. 

\section{Discussion}

Ergodic chaos in long-term evolutionary dynamics offers two main conceptual perspectives.
First, frequency-dependent ecological interactions can generate complicated evolutionary trajectories that visit all feasible regions of phenotype space in the long run even if the external environment (given by system parameters) is constant. In such situations, the current phenotypic state of a population can never be understood as the result of an equilibrium or optimisation process, even though the process determining the phenotypic state is entirely adaptive and deterministic.  

Second, chaotic evolutionary trajectories are intrinsically unpredictable. At the very basic level, biological evolution is stochastic, because the single-molecule events that correspond to
spontaneous mutations are subject to fundamental, quantum mechanical randomness.  The adaptive dynamics models considered here are defined on a coarse-grained time scale that is much larger than the typical time between mutations, and the models generate deterministic trajectories that are shaped by ecological interactions. However, mutations can in principle fundamentally alter chaotic long-term evolutionary trajectories due to 
sensitive dependence on initial conditions: whether or not a
particular mutation occurs at a given point in time may slightly
change the initial condition for the evolutionary trajectory unfolding
after that time point, and this difference may translate into vastly
different phenotypic states at much later points in time. It is
important to note, however, that when the adaptive dynamics has a
positive Lyapunov exponent and hence shows divergence along some
composite direction in a high-dimensional phenotype space, divergence
between trajectories in any given phenotypic component may still take
a long time  (Fig. 5). This is because even if the largest
Lyapunov exponent is positive, the total number of positive Lyapunov exponents usually
represents a small fraction of the dimensionality of phenotype space.
Therefore, the probability that directions with positive Lyapunov exponents have a significant component in two randomly chosen dimensions is small, which explains the initial lack of divergence in the two projected trajectories shown in Fig.~5. On the other hand, the probability that a projection is exactly 0 in all directions with positive Lyapunov exponents is 0, and hence divergence will eventually occur. Extrapolating from this, replaying the tape of life \cite{gould1989} could potentially result in completely different outcomes if evolution is chaotic.

Our results are also relevant for the general problem of the prevalence of chaos in multidimensional dynamical
systems. Chaos is well studied in high-dimensional Hamiltonian systems \cite{zaslavsky_etal1991}, as well as in discrete-time systems of coupled oscillators with many degrees of freedom \cite{kaneko1989,ishihara_kaneko2005}, but these results are not applicable to dissipative, non-Hamiltonian systems in continuous time, such as the adaptive dynamics models studied here. Surprisingly, our analysis may be the first systematic study of the occurrence of chaos in such systems as a function of the dimensionality of phase space. 

Because the likelihood of chaotic evolutionary dynamics in our models is strongly influenced by the dimensionality of phenotype space, the biological relevance of our results hinges on the number of phenotypic properties affecting ecological interactions in real systems, and on the potential for epistasis between these phenotypic properties. Given that metabolic networks of even the simplest bacterial organisms such as {\it E. coli} are incredibly complex \cite{selkov_etal1996,karp_etal1999}, it seems likely that in general, many phenotypic properties combine in complicated ways to affect ecological interactions such as competition for resources. Data testing this directly seems to be scant, but an indication for high-dimensionality of ecologically relevant phenotype space comes for example from studies of the genetics of adaptive diversification in fishes \cite{keller_etal2012} and in bacteria \cite{herron_doebeli2012}. In pairs of fish species that have recently speciated into two different ecotypes, there is abundant genetic differentiation between the species, which is distributed over the whole genome. Some of this differentiation is due to differences in mating preferences, but it appears that many of the observed genetic differences are related to ecological traits, and hence that many different genes affect ecological properties, and thus ecological interactions, of these fish \cite{keller_etal2012}. Similarly, a genetic analysis of adaptive diversification in evolution experiments with {\it E. coli} revealed that the diversified ecotypes that evolved from a single ancestral strain in ca. 1000 generations differed in many genes carrying adaptive mutations \cite{herron_doebeli2012}. Other recent evolution experiments with  {\it E. coli} showed the importance of epistasis for evolutionary dynamics \cite{tenaillon_etal2012}. It would seem to be an important empirical endeavour to gain a general understanding of the number of different phenotypic properties that can be expected to affect ecological interactions, and of the degree of epistasis between them.

Even if one accepts the premise of high-dimensional phenotype spaces, one could question the realism of the logistic competition models used here. While it is true that our models do not derive from an underlying mechanistic model for ecological interactions between individual organisms, our statistical approach examines in some sense all possible competition models whose trajectories are confined to a bounded region in phenotype space. This is because, to second order, any such model for the evolutionary dynamics of $d$ phenotypic components will have the general form (6). In particular, the subset of ``realistic'' models will have this form. Thus, if essentially all models of the form (6) have chaotic dynamics for high $d$, then any particular realistic model is likely to have such dynamics as well.

For now, our results warrant at least a critical re-examination of the generality of simple equilibrium and optimization dynamics in evolution. 40 years ago, the realization that simple ecological models can have very complicated dynamics revolutionized ecological thinking \cite{may1976}. Our high-dimensional models are not simple, but they show that non-linear evolutionary feedbacks generated by frequency-dependent ecological interactions can also lead to very complicated dynamics. In fact, with frequency dependence most evolutionary dynamics may be chaotic when phenotypes are high-dimensional. In general, evolution is a complicated dynamical systems driven by birth and death events that are determined by many different factors, such as external biotic and abiotic conditions, current phenotype distributions, age and physiological condition, etc. If birth and death rates are complicated functions of many different factors that change themselves as evolution unfolds, we do not see any reason to expect that in general, evolutionary dynamics should be simple (after all, it is for example well known that weather often exhibits chaos and long-term unpredictability, essentially because of the nonlinearity of the dynamics and the complexity of the interactions between the many different components determining the weather). 
Nevertheless, our perception is that to date, evolutionary biologists are unaware of the fact that general evolutionary dynamics in continuous phenotype spaces of high dimensions are likely to be very complicated. Knowing that, in principle, long-term evolutionary complexity can be due to intrinsic frequency-dependent interactions rather than simply to changes in the external environment would generally seem to be useful, in the same way as it was useful when, four decades ago, ecologists became aware of the possibility of chaos due to non-linear interactions in generic ecological models. Our results indicate that chaos and complexity in long-term evolutionary dynamics should be given serious consideration in future studies.

\section{Methods}

Here we describe how the largest Lyapunov exponent is calculated, illustrate the divergence of chaotic trajectories, and provide approximate analytical explanations for the numerical results
reported for the size of chaotic attractors (Figure 2), the probability of chaos as a function of the dimension $d$ of phenotype space (Figure 3), and the scaling of the largest Lyapunov exponent with $d$ (Figure 4). 

\subsection{Calculation of Lyapunov exponents}
For each trajectory obtained through numerical
integration of the adaptive dynamics (6), the time average of the largest Lyapunov exponent $\l $ was calculated as follows. Every $\t$ time
units the trajectory was slightly perturbed, $x'=x+\d x_0$, by a vector
with a constant magnitude $ \| \d x_0 \|$ and a random direction. Both
resulting trajectories were propagated for $\t$ time units, after which the
distance between the perturbed and unperturbed positions  $ \| \d x_{\t} \|$ was recorded. The largest
Lyapunov exponent was calculated for each $\t$ time units as
\begin{align}
\label{ll}
\l = \frac {1}{\t} \ln \left(\frac {\| \d x_{\t} \|}{\| \d x_{0} \|}\right),
\end{align}
and subsequently averaged over the trajectory.
Visual inspection of trajectories led us to the following selection
criteria: Trajectories with the $ \l  > 0.1$ were
usually chaotic, trajectories with the $| \l | < 0.1$ were
quasi-periodic, and trajectories with  $\l < - 0.1$
converged to fixed points.  

Choosing suitable time intervals $\t$ for the numerical calculations of the largest Lyapunov exponent is constrained on both
sides. On the one hand, values of $\t$ that are too small do not leave sufficient time for a randomly chosen
direction of perturbation to align itself with the direction of the
fastest divergence corresponding to the largest Lyapunov exponent. On the other hand,
for values of $\t$ that are too large the divergent
trajectories reach the limit of the available phenotype space and fold
back, so that the distance between them saturates.  
Both scenarios result in the underestimation of the distance
between trajectories and the resulting value of the largest Lyapunov
exponent.  Thus the optimal value of $\t$ is the one
giving the maximum average value of the largest Lyapunov
exponent. Our numerical experiments indicate that the optimal values
of $\t$ lie in the range between $10^{-5}$ and $10^{-3}$, with the
smaller values better suited for higher dimensions $d$.  

\subsection{Divergence of evolutionary trajectories}
Consider two trajectories initially separated by a
small distance, say $ \| \d x_0 \|=10^{-3}$. Evolution of both trajectories is given by the adaptive dynamics (6). For high dimensions of phenotype space ($d=150$ in
this example), the adaptive dynamics is almost certainly chaotic, so one
would expect an exponential growth of the distance between
trajectories. In Fig.~5 we show an example of such
divergence. Contrary to naive expectations, the visual divergence
commences not immediately, but only after a noticeable transient period
during which two trajectories remain essentially indistinguishable. 
In reality,
the two trajectories start to diverge immediately, but this divergence often remains invisible in a two-dimensional
projection of 150-dimensional phenotype space, such as that shown in Fig.~5.
This occurs because even if the largest
Lyapunov exponent is positive for a high-dimensional
system, the total number of positive Lyapunov exponents usually
represents a small fraction of the dimensionality of the space. Therefore, the probability that directions with positive Lyapunov exponents have a significant component in two randomly chosen dimensions is small. This explains the initial lack of divergence in the two projected trajectories shown in Fig.~5. On the other hand, the probability that a projection is exactly 0 in all directions with positive Lyapunov exponents is 0, and hence eventually the projected trajectories will indeed diverge, as shown in Fig.~5. Thereafter, the trajectories usually evolve completely independently.
\begin{figure}[b]
\includegraphics[width=.45\textwidth]{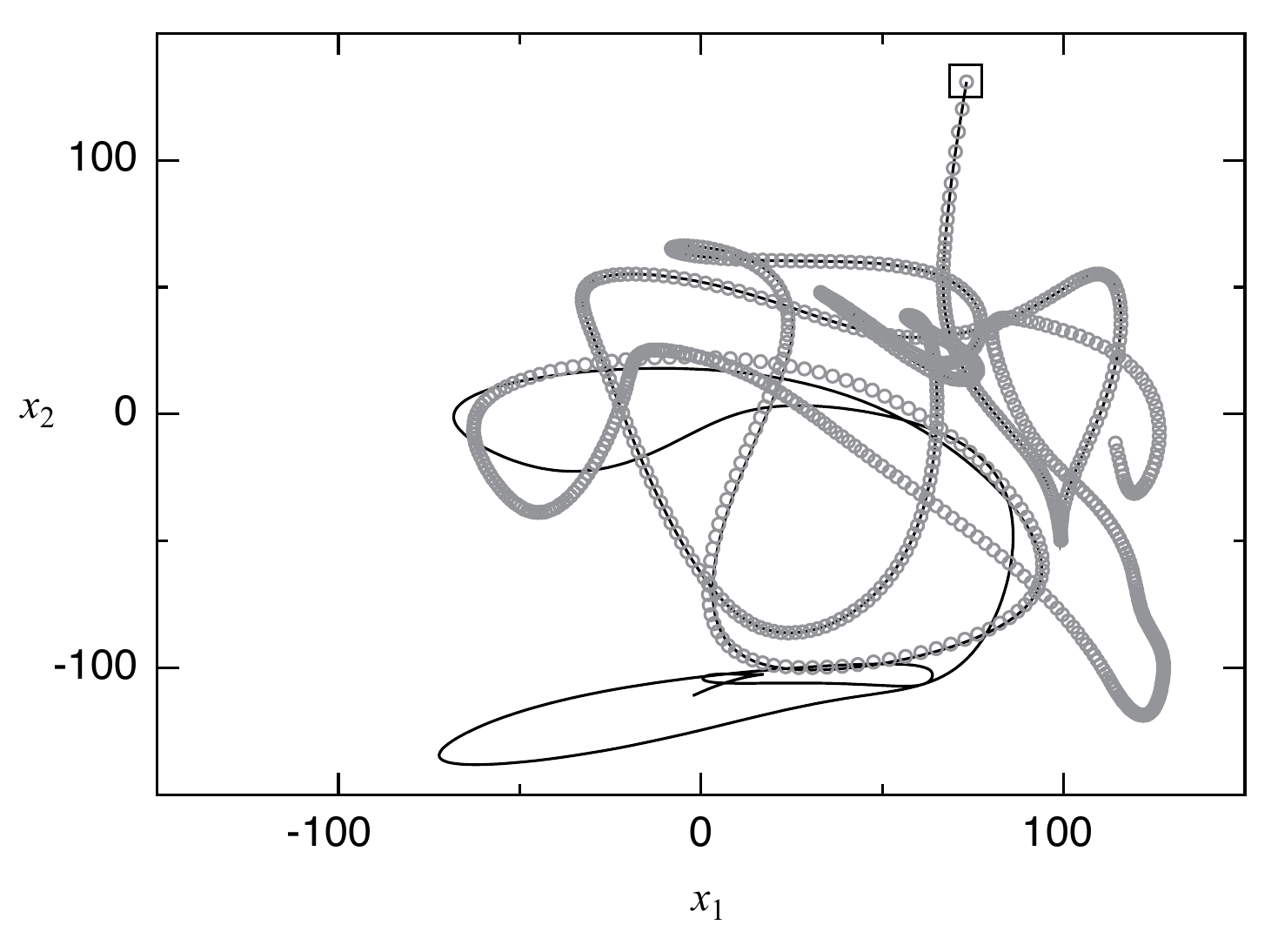}
\caption{\label{f5} Example of divergence of two evolutionary trajectories (black
  line and red circles) in
  $d=150$-dimensional phenotype space. The trajectories start from initial conditions separated by a visually undetectable small distance $\| \d
  x_0 \|=10^{-3}$. The initial positions of both trajectories are marked by
  a square at the top of the plot. The total time of evolution of both
  trajectories is $100/d^2\approx 0.0044$. The spacing of the circles indicates the speed with which the trajectory unfolds. The two trajectories stay close for a long time but eventually diverge and move to completely different regions in phenotype space.
}
\end{figure}

\subsection{Size of chaotic attractors and magnitude of Lyapunov exponents.}
First we consider the scaling of the spatial 
coordinates, $x_i\sim d$, illustrated in Fig.~3 and in Figs.~6,~7  and the scaling of the Lyapunov
exponent, $\l \sim d^2$  (Fig.~4). 
\begin{figure}
\includegraphics[width=.45\textwidth]{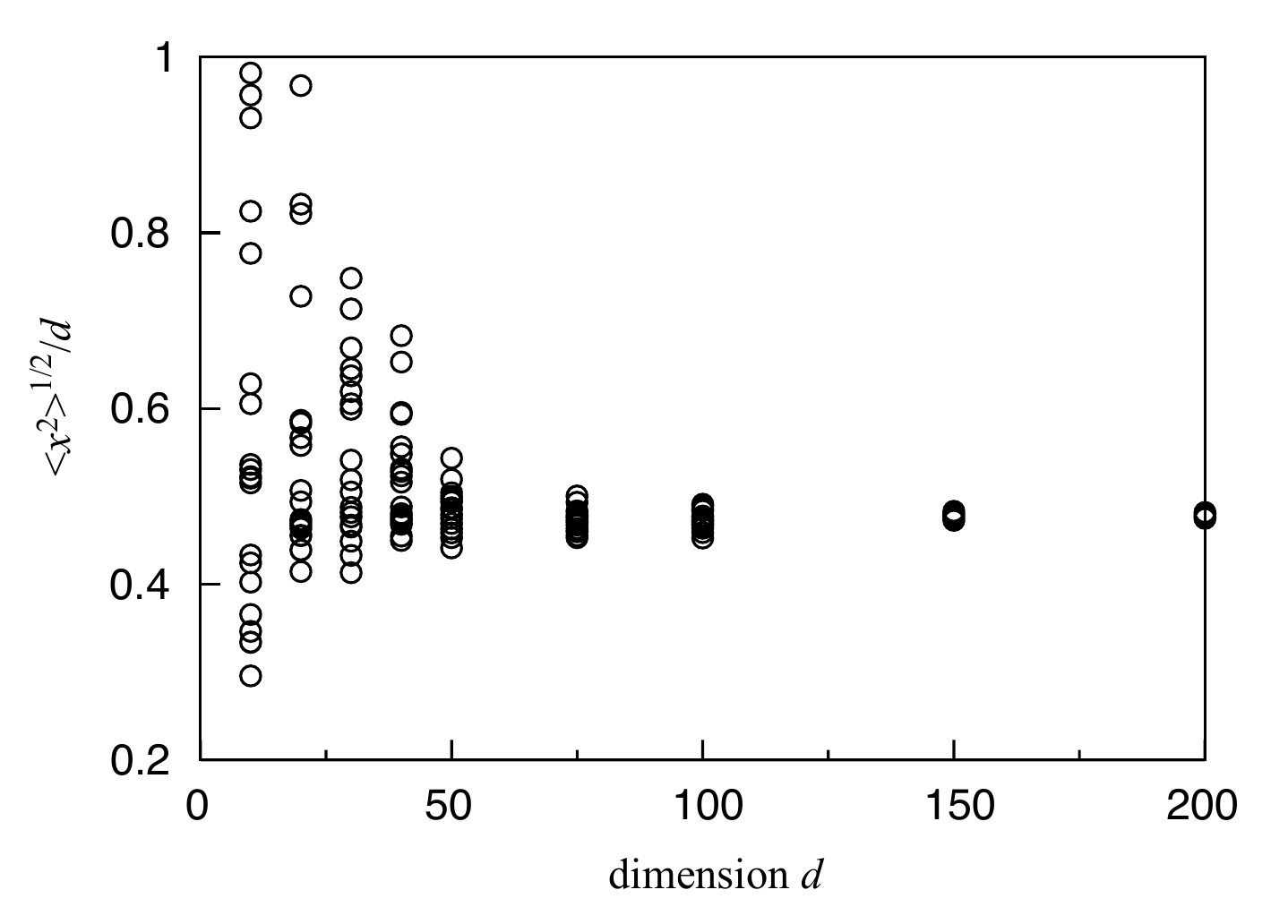}
\caption{\label{f6}Scaling of the mean square of a coordinate $x_i$
  with the dimensionality of phenotype space,
  $\sqrt{\langle x_i^2 \rangle} \sim d$. Same data set as used for Fig. 2.}
\end{figure}
\begin{figure}
\includegraphics[width=.45\textwidth]{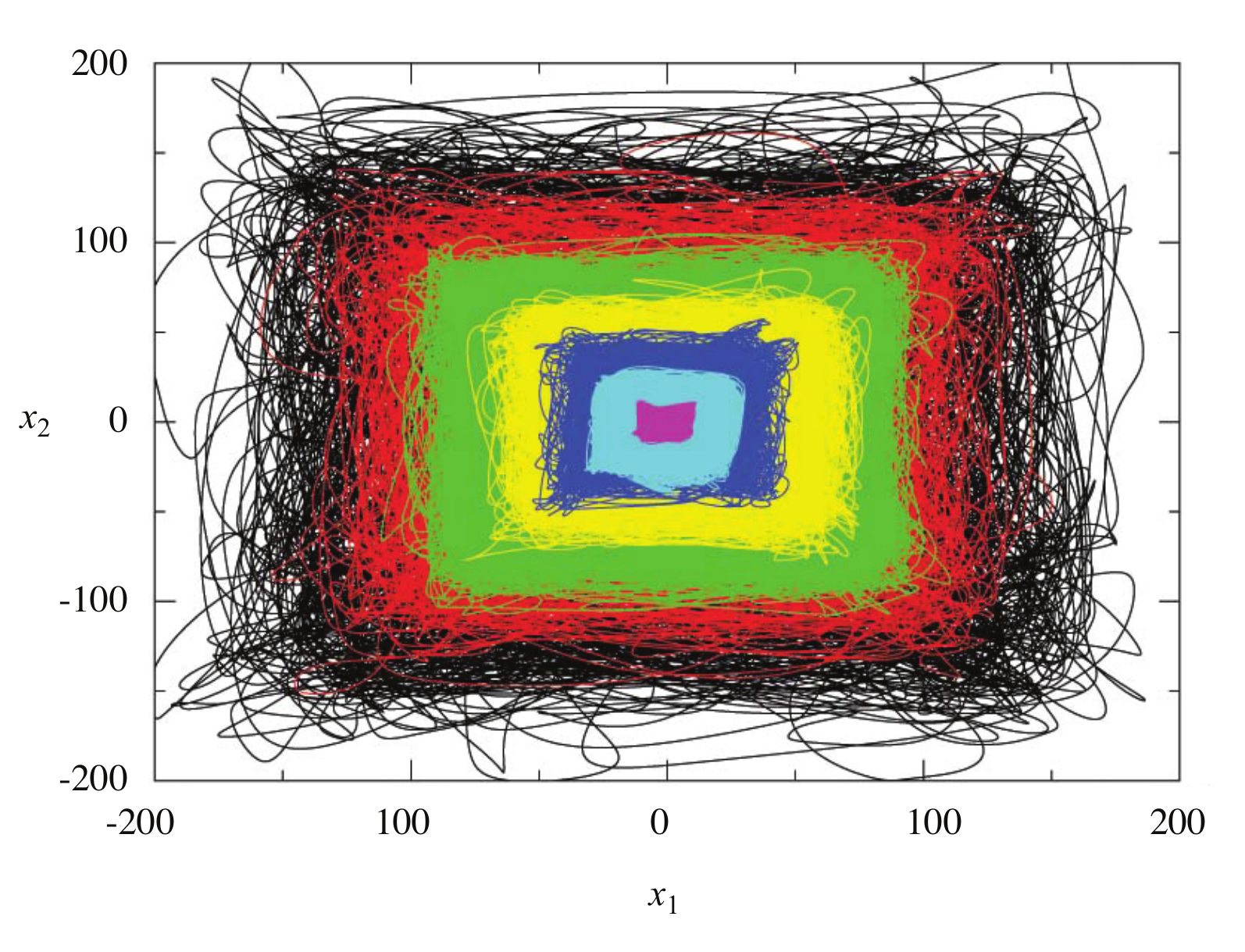}
\caption{\label{f7} Examples of projections of ergodic chaotic trajectories for 
$d=15$ (magenta), $d=30$ (blue), $d=50$ (dark blue), $d=75$ (yellow),
$d=100$ (green), $d=150$ (red), and $d=200$ (black). The figure illustrates the scaling $x_i\sim d$. }
\end{figure}
If we make the reasonable assumption that each phenotypic coordinate has  a similar
scale, $x_i \sim x$, the dynamical system (6) becomes
\begin{align}
\label{scale0}
\frac{dx}{dt}= x \sum_{j=1}^d b_{ij}  + x^2 \sum_{j,k=1}^d a_{ijk}- x^3
\end{align}
for $i=1,...,d$. Here the $b_{ij}$ and $a_{ijk}$ are identically distributed random
terms with zero mean and unit variance, and a typical value of the sum of $d$ such terms is the standard deviation $\sqrt{d}$, which yields
\begin{align}
 \label{scale}
\frac{dx}{dt}=\sqrt{d} x  +d x^2 - x^3.
\end{align}
Introducing new variables,
\begin{align}
\label{scale_v}
y&=\frac{x}{d} \\
\nonumber
\theta &= t d^2,
\end{align}
we convert (\ref{scale}) into the differential equation 
\begin{align}
\label{scale_eq}
\frac{dy}{d\theta}=\frac{y}{d^{3/2}}  + y ^2 - y^3
\end{align}
with two universal, $d$-independent terms and a linear term that
vanishes in the limit of large $d \gg 1$. 
The transformation (\ref{scale_v}) explains the observed scaling of
the size of chaotic attractors, $x=yd$ (Figs.~3,~6,~7), and of the largest Lyapunov
exponent, whose dimension is the inverse of time, $1/t=d^2/\theta$ (Fig.~ 4). The transformation also shows  that the linear term 
$\sum_{j=1}^d b_{ij} x_j$ in (6) does not produce any significant effect on
the probability of chaos and on the form of the attractor for large $d$. 

 Taking into account  (\ref{scale},\ref{scale_v}), it is
possible to rescale the coefficients $b_{ij}$ and $a_{ijk}$ in such
a way that the total ``strength'' of epistatic interactions between the phenotypic components, as well as the size of the area of phenotype space filled out by ergodic trajectories,
do not depend on the dimension $d$ of phenotype
space.  Eq.~(6) with $b_{ij}$ and $a_{ijk}$ drawn from Gaussian
distributions with variances $1/d$ and $1/d^2$ (so that their typical
values are $1/\sqrt{d}$ and $1/d$) produces similar-sized ergodic
attractors for various $d$, as shown in Fig.~8. A slight asymmetry of the
trajectories in Fig.~8 is caused by the linear terms  $b_{ij}$ which
are no longer irrelevant. Without these terms the high-$d$ trajectories of the
rescaled Eq.~(6) are completely analogous to the trajectories of the original Eq.~(6) when the latter are rescaled through division by
$d$. 
\begin{figure}
\includegraphics[width=.45\textwidth]{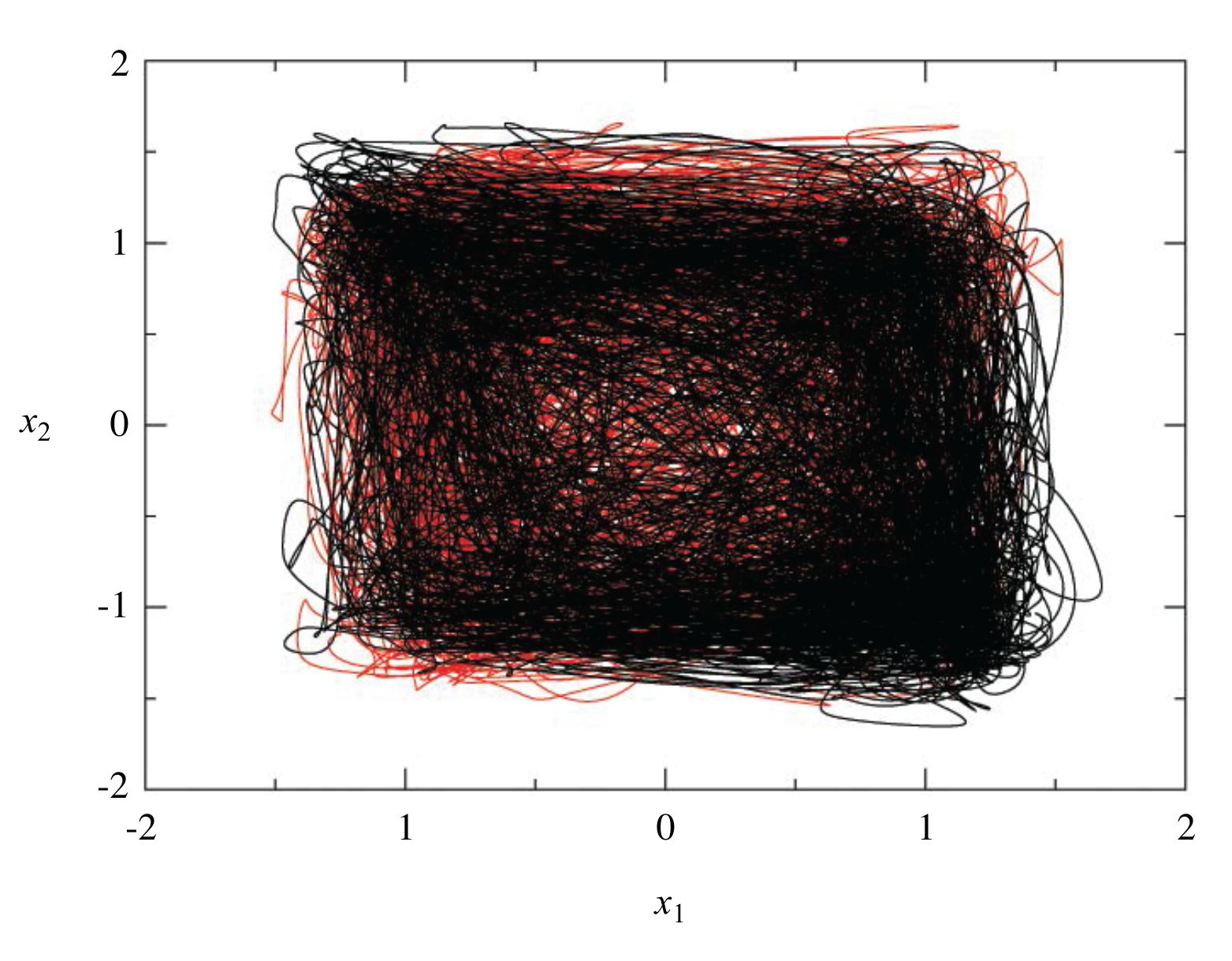}
\caption{\label{f8} Examples of projections of ergodic chaotic
  trajectories system (6) (main text) when the coefficients  $b_{ij}$ and $a_{ijk}$ were divided
  by $\sqrt{d}$ and $d$, respectively; 
$d=75$ (red), and $d=100$ (black).}
\end{figure}

\subsection{Probability of chaos.}
Second, we provide an explanation for the increase in the occurrence of chaos with the dimension $d$ of phenotype space.
Stationary points of the adaptive dynamics (6) are defined as
solutions of the corresponding system of algebraic equations where the
right-hand side is set equal to zero. Generally, a
system of $d$ third-order algebraic equations has $3d$ solutions (some of which
may coincide), and hence the dynamical system (6) has $3d$ stationary points.  We
propose that the system is chaotic if all these stationary points are
unstable in at least one direction, i.e., if at least one
eigenvalue of the local Jacobian matrix $J$ at each stationary point
$x^*$ has a
positive real part. If $P_{n}$ is the probability that the real
part of an eigenvalue is negative, and assuming that all Jacobian eigenvalues are
statistically independent, the probability that at least one out of
$d$ eigenvalues of the Jacobian at a stationary point has a positive real part
is $1-P_{n}^d$. Hence the probability of chaos is 
\begin{align}
\label{P_chaos}
P_{chaos}=(1-P_{n}^d)^{3d}.
\end{align}
If $x^*$ is a stationary point of (6), the elements of the Jacobian matrix $J(x^*)=\left(J_{ij}(x^*)\right)_{i,j=1}^d$ consist of two terms, 
 \begin{align}
\label{jacobian}
J_{ij}(x^*)&=\sum_{k=1}^d
(a_{ijk}+ a_{ikj})  x_k^* - 3 x_i^{*2} \d_{ij}\\
\nonumber
&=J_{ij}^{(1)}+J_{ij}^{(2)},
\end{align}
where $\left(\d_{ij}\right)$ is the identity matrix.
Here we ignored  the linear term $\sum_{j=1}^d b_{ij} x_j$
which we argued above to be negligible for increasing $d$. 
We assume that  the distribution of $x^*_i$  is the same as
for the coordinates $x_i$ themselves and is given by the universal
invariant measure shown in Fig.~ 3. This assumption allows us to consider the two terms $J_{ij}^{(1)}$ and $J_{ij}^{(2)}$ as statistically independent. The first
term, $J_{ij}^{(1)}=\sum_{k=1}^d (a_{ijk}+ a_{ikj})  x_k^*$, is a sum of a large
number $d$ of random variables with zero mean and a finite variance. 
 Note that this is true for any distribution of $a_{ijk}$ which decays sufficiently fast
  with $|a_{ijk}|$, and hence the results reported here are true for any distribution of the coefficients $b_{ij}$ and $a_{ijk}$ with a finite variance.  According to
the Central Limit Theorem, this sum is a Gaussian-distributed variable with zero
mean and variance $\a^2 = 2d \langle x^2 \rangle$. 
``Girko's circular law'' \cite{girko} states that eigenvalues of a random 
$d \times d$-matrix with Gaussian-distributed elements with zero mean and unit variance are uniformly
distributed on a disk in the complex plane with radius $\sqrt{d}$. Thus, the eigenvalues of
$J_{ij}^{(1)}$ are  uniformly distributed on a disk with radius
$d\sqrt {2\langle x^2 \rangle}$. The
probability for an eigenvalue to have real part $rd\sqrt
{2\langle x^2 \rangle}$,  with $|r| \leq 1$, is then
proportional to the length of the chord intersecting the radius of the disk at the point
$r$,
\begin{align}
\label{eigen}
P_c(r)=\frac{2\sqrt{1-r^2}}{\pi}.
\end{align}
(The factor $2/\pi$ normalizes the distribution to one.)
 Considering  the second, diagonal,  term of the Jacobian, $J_{ij}^{(2)}=- 3
x_i^{*2} \d_{ij}$,  we rely on the numerical observation that the distribution function $P(x)$
 has a universal form, which is independent of $i$, and whose scaled form $P(y)$, with $y=x/d$, is given in Fig.~3. 

Both $J_{ij}^{(1)}$ and $J_{ij}^{(2)}$ contribute terms of order $d^2$ to the eigenvalues of the Jacobian. The contribution from $J_{ij}^{(1)}$ may have a positive or a negative real part, and the probability that it has a negative real part is $1/2$. The contribution from $J_{ij}^{(2)}$ is always negative and has magnitude $-3x^2$ with probability $P(x)dx$. If the contribution of $J_{ij}^{(2)}$ is $-3x^2$, the probability that the sum of the two contributions has negative real part is $ \int_0^{3x^2/\a} P_c(r)dr$. If we rescale everything by $d^2$, we thus obtain the the probability that the real
part of an eigenvalue of the Jacobian is negative as
\begin{align}
\label{ps1}
P_{n}=\frac{1}{2} + \int_{-\infty}^{+\infty} P(y) dy \int_0^{3y^2/\b} P_c(r)dr.
\end{align}
Here the 1/2 term reflects the probability that the eigenvalue of 
$J_{ij}^{(1)}$ has a negative real part, $\b=\a/d^2=\sqrt{2 \langle
  y^2 \rangle}$, and the double integral gives
the probability that the positive real part of $J_{ij}^{(1)}$ is
smaller than the  contribution $-3 y^2$ from $J_{ij}^{(2)}$.
Integration on $dr$ produces
\begin{align}
\label{ps2}
P_{n}=\frac{1}{2} \left[ 1 + \int_{|y|>\sqrt{\b/3}} P(y) dy \right] \\
\nonumber
+\int_{|y|<\sqrt{\b/3}} 
\frac{\sin^{-1}(3y^2/\b) + 3y^2/\b \sqrt{1-(3y^2/\b)^2}}{\pi}P(y) dy.
\end{align}
Using the numerical data for $P(y)$ shown in Fig.~3, we calculate $\b
\approx 0.675$
and perform numerical integration of $P(y)$ to obtain $P_{n} \approx 0.85$. Substituting this value into Eq.~(\ref{ps2}) above provides a reasonable fit for the
observed probability of chaos, as illustrated in Fig.~9.
\begin{figure}
\includegraphics[width=.45\textwidth]{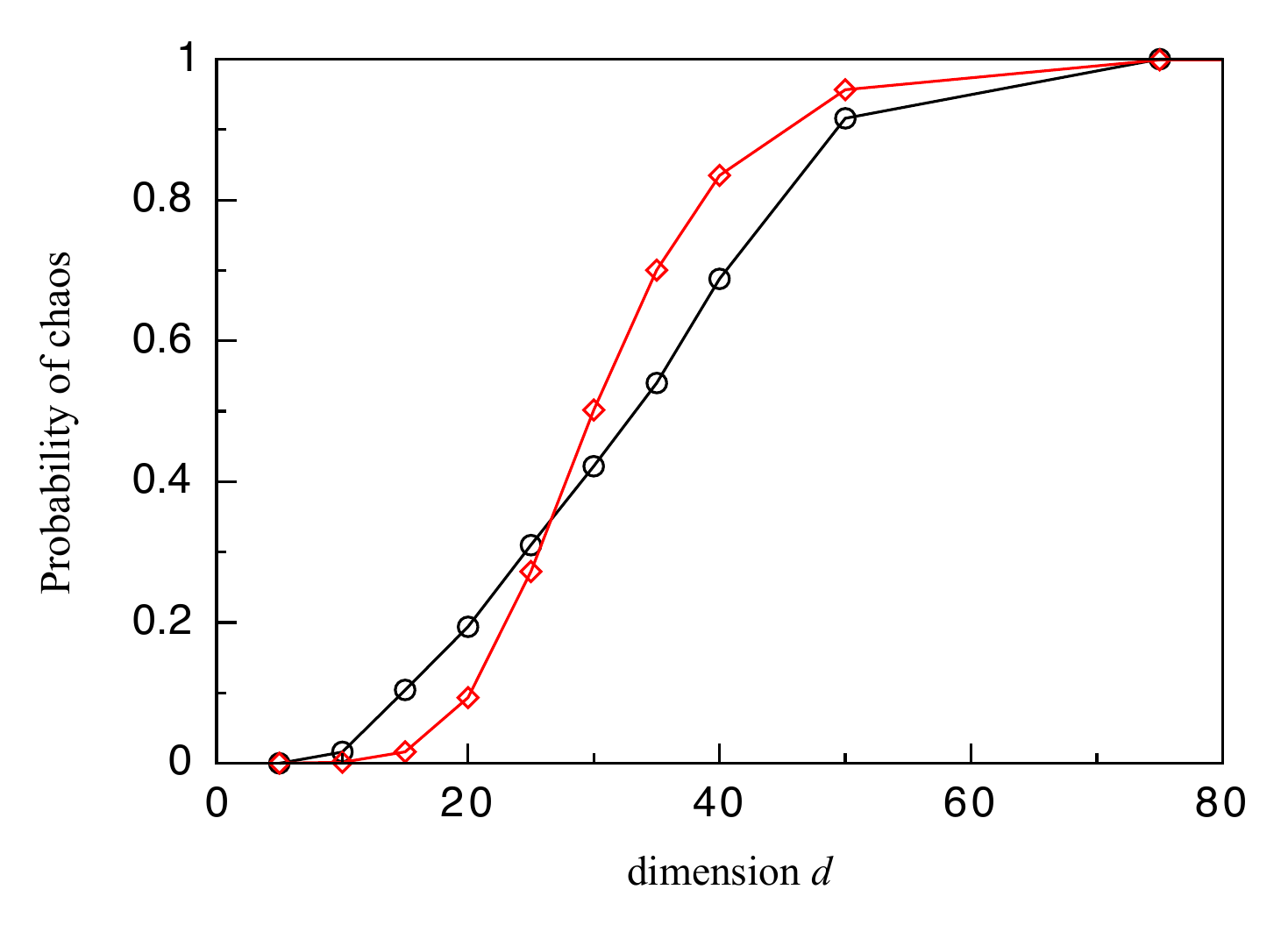}
\caption{\label{f9} The numerically measured probability of occurrence
 of chaos (black line; data from Fig. 2)  and the estimate (\ref{P_chaos})
 with $P_{n} = 0.85$ (red line), which shows a reasonable fit. We note that formula
 (\ref{P_chaos}) is rather sensitive 
to the value
of $P_n$, with variations in $P_n$ by as little as 0.01 leading to noticeably different plots.  
}
\end{figure}
It is important to note that while the details of the calculations of $P_n$
depend on the particular form of the dynamical system (6) and its
Jacobian matrix, the conclusion that the probability of chaos increases with the dimension $d$ is general: If
each eigenvalue has a non-vanishing probability to have a positive
real part, $P_n<1$,  the probability of chaos given by (\ref{ps2}) approaches one for $d \rightarrow \infty$.

\subsection{Scaling of Lyapunov exponents.}
\begin{figure}
\includegraphics[width=.45\textwidth]{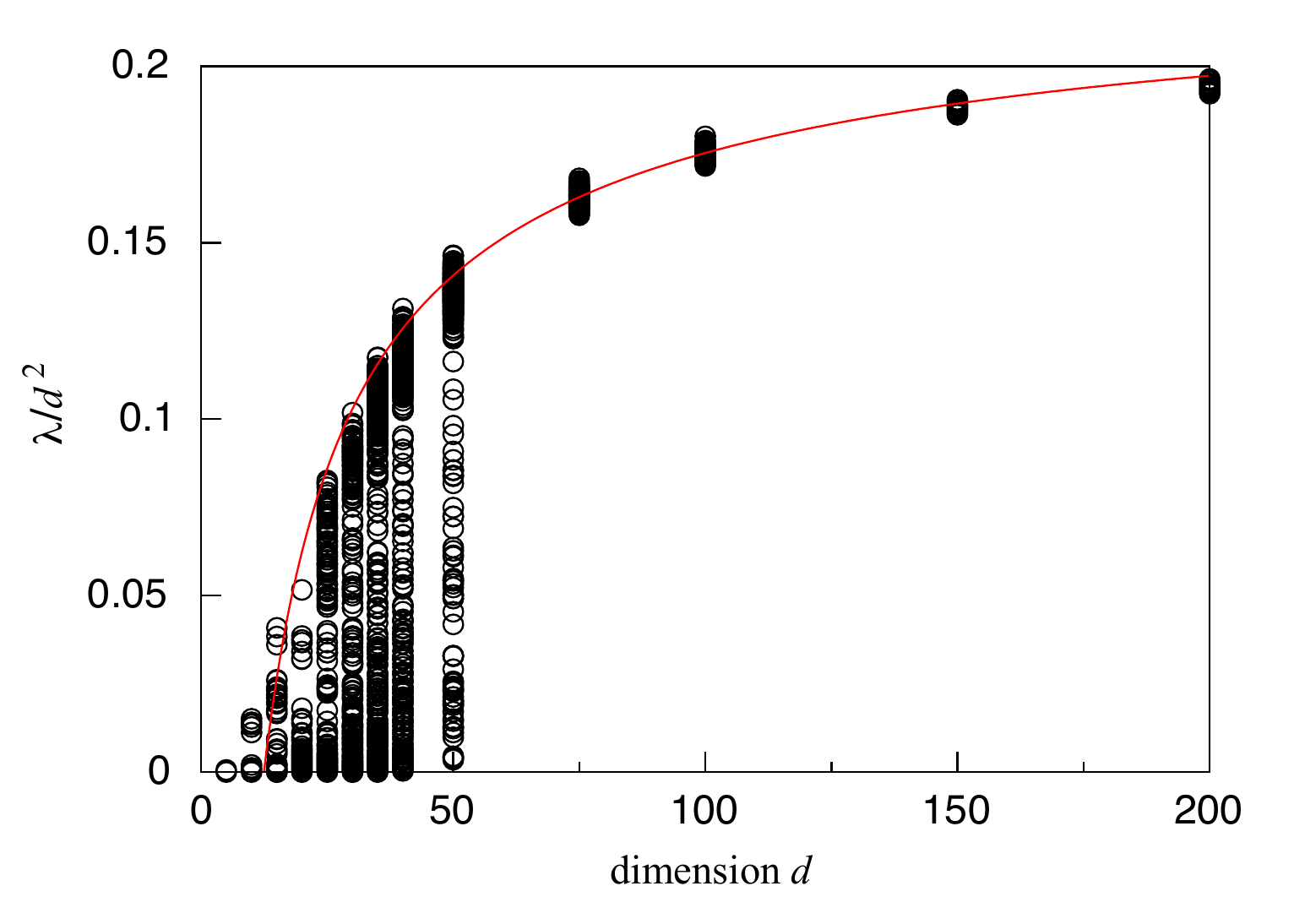}
\caption{\label{f10} The slow convergence to the asymptotic regime of the average largest Lyapunov
 exponent $\l$ (black circles; same as Fig. 4) is explained using the extreme value statistics given by Eq. (\ref{extr3}) (red line).
}
\end{figure}

Finally, we show that  the slow convergence of the largest Lyapunov
exponent $\l$ to its scaling asymptotic, as shown in Fig. ~4, can be
explained as a general consequence of extreme value statistics. 
We again use the arguments provided above for the fact that the 
eigenvalues of $J^{(1)}$ are distributed according to ``Girko's
circular law''. Given the distribution $P_c(r)$ of real
parts of rescaled eigenvalues of $J^{(1)} $, we calculate the distribution of the
largest real part of the rescaled eigenvalue:
\begin{align}
\label{extr1}
P_{max}(\l) = P_c(\l) d \left[ \int_{-1}^{\l}P_c(r) dr \right ]^{d-1}.
\end{align}
Here the term $ P_c(\l)$ is the probability that the largest
eigenvalue is equal to $\l$, and the integral term gives the probability that
the remaining $d-1$ eigenvalues are less than $\l$. The factor $d$
reflects the fact that any of $d$ eigenvalues could be the largest. 
To calculate the average value of the largest eigenvalue, 
\begin{align}
\label{extr2}
\langle \l \rangle = \int \l P_{max}(\l) d\l,
\end{align}
we substitute (\ref{eigen}) into (\ref{extr1}) and integrate
(\ref{extr2}) by parts, obtaining
\begin{align}
\label{extr3}
\langle \l \rangle = \l^* - \int_{-1}^{1} \left[ \frac{\arcsin(x) +
    \pi/2 + x\sqrt{1-x^2}}{\pi} \right]^d dx,
\end{align}
where $\l^*$ is a constant representing the upper limit of the
rescaled largest eigenvalue. Above we ignored the scaling coefficient
for $\l$ and the contribution of the diagonal part of Jacobian
$J_{ij}^{(2)}$, thus were unable to explain the numerical value for this upper limit, $\l^* \approx  0.235$.
However, our simple estimate based on the extreme value
statistics provides a reasonable description of how the largest eigenvalue $\l$
approaches its asymptotic value $\l^*$ as $d\rightarrow \infty$, Fig.~\ref{f10}.

\begin{acknowledgments}
I. I. was supported by FONDECYT (Chile) project \#1110288. M. D. was supported by NSERC (Canada). Both authors contributed equally to this work. 
\end{acknowledgments}

\bibliography{chaos}
\bibliographystyle{prslb}


\end{document}